\documentclass[twocolumn,trackchanges]{aastex6}

\begin{document}
\title{Characterizing the stellar population of NGC 1980}

\author{Marina Kounkel\altaffilmark{1}, Lee Hartmann\altaffilmark{1}, Nuria Calvet\altaffilmark{1}, Tom Megeath\altaffilmark{2}}
\altaffiltext{1}{Department of Astronomy, University of Michigan, 1085 S. University st., Ann Arbor, MI
48109, USA}
\altaffiltext{2}{Ritter Astrophsical Research Center, Department of Physics and Astronomy, University of Toledo, Toledo, OH 43606, USA}
\email{mkounkel@umich.edu}

\begin{abstract}
NGC 1980 is a young cluster that is located about 0.5 degrees south of the Orion Nebula Cluster (ONC). Recent studies by Bouy et al. and Pillitteri et al. have suggested that NGC 1980 contains an older population of stars compared to a much younger ONC, and that it belongs to a foreground population that may be located in front of the Orion A molecular gas by as much as 40 pc. In this work we present low-resolution spectra towards 148 young stars found towards the NGC 1980 region. We determine the spectral types of these stars, examine accretion signatures and measure the extinction towards them. We determine that based on these observations, the age of the population of NGC 1980 is indistinguishable from L1641, estimated to be $\sim$3 Myr, comparable with the study by Fang et al.
\end{abstract}


\section{Introduction}

The Orion A molecular cloud contains the nearest regions of massive star formation. Its most active component is the Orion Nebula Cluster (ONC), which contains a few thousands of young stellar objects (YSOs) ranging in mass from hydrogen burning limit to massive O stars. Extending to the south of the ONC is the L1641 cloud, which contains a somewhat more distributed population of YSOs. NGC 1980 is a cluster found between these two stellar populations. Loosely associated wtih a massive O9 binary star, $\iota$ Ori, this cluster could be an important jigsaw puzzle piece in understanding the star formation history of the region.

NGC 1980 is thought to be an older cluster than the population of stars that belong to the ONC \citep[$<$1--3, Myr][]{2010dario} and L1641 \citep[$\sim$ 3 Myr,][]{2012Hsu}. \citet{2012alves} have estimated its age to be 4-5 Myr from the lifetime of $\iota$ Ori, and later revised the estimate to 5--10 Myr based on the optical colors of the stars in the vicinity of the region \citep[hereafter BA14]{2014bouy}. \citet{2013Pillitteri} performed an X-ray survey of Orion A; they found that towards NGC 1980 there exists an excess of low-extinction sources with A$_V\sim$0.1 compared to L1641, and that it also contained a one of the largest fraction of YSOs that have already lost their dusty disks, with an age estimate of $\geq$5 Myr. More recently, however, \citet{2017fang} suggested that the age of NGC 1980 is closer in agreement to that of the ONC, with the median age of 1.8 Myr.

\citet{2012alves} and BA14 have questioned the degree to which NGC 1980 is related to the ONC, claiming it to be a foreground population of stars (by as much as 10--40 pc) located at $\sim$380 pc. \citet{2013Pillitteri} have also argued in support of it, demonstrating that the X-ray luminosities are 0.3$\pm$0.1 dex higher in the northern end of L1641 (onto which NGC 1980 is superimposed) compared to the southern end, suggesting that NGC 1980 is 1.4 times closer than the main cloud.  However, recently \citet{2017kounkel} have found a distance toward the ONC of 388$\pm$5 pc using stellar parallaxes of non-thermally emitting YSOs; this is a revision to the previously commonly accepted distance of 414$\pm$ 7 pc \citep{2007menten}. In addition, \citet{2017kounkel} found the distance towards the southern end of L1641 of 428$\pm$10 pc; the difference in X-ray luminosities is most likely caused solely due to the spatial orientation of the cloud. Additionally, radial velocity monitoring of the ONC by \citet{2016kounkela} and \citet{2016dario} showed that the radial velocities of the stars towards NGC 1980 are indistinguishable from those of the molecular cloud; in fact, this agreement is better than towards any other region in the ONC. A possible argument in favor of NGC 1980 belonging to a foreground population may persist due to the lack of any reflection nebulosity towards $\iota$ Ori. However, the degree to which the the $\iota$ Ori is related to the cluster is unclear, particularly since this was involved in a dynamical ejection event \citep{2001deZeeuw, 2004Gualandris}. With these updated distances and radial velocity measurements, the nature of NGC 1980 as a foreground cluster falls into a considerable doubt. 

NGC 1980 has been left out in many surveys of star formation in Orion that focused on the properties of individual stars. It is located southward of the Trapezium cluster, the heart of the star formation in the ONC, and it is outside of the field-of-view of the studies by \citet{1997hillenbrand} and \citet{2010dario}. On the other hand, it falls northward of the region covered by the surveys of L1641 filament \citep{2009Fang,2012Hsu,2013Hsu}. Similarly to the work by \citet{2017fang}, in this paper we seek to bridge the gap between these works. We present an analysis of the low resolution optical spectra of stars identified by BA14 towards this cluster. We asses the youth and determine the spectral types of these stars, as well as measure the equivalent width of H$\alpha$ line (Section \ref{sec:data}). On the basis of these measurements, we classify these as sources as Classical T Tauri Stars (CTTS) and Weak-lined T Tauri Stars (WTTS) based on the definitions from \citet{2003White}. Through analyzing the ratio of CTTS to WTTS, we compare the age of NGC 1980 to that of the ONC and L1641 to confirm its older age (Section \ref{sec:acc}). Finally, we estimate the extinction to these stars (Section \ref{sec:red})

\section{Data}\label{sec:data}

We conducted observations using the IMACS spectrograph on the Magellan Baade telescope \citep{imacs}. Observations were done using the $f/2$ camera with the 300 line grism at a blaze angle of 17.5$^\circ$ and slit width of 0.6''. In this configuration, the typical resolution is 4 \AA, and the spectral coverage is approximately 4000--9000 \AA. However, this coverage may be truncated for the stars close to the edge of the field. Additionally there is a chip gap $\sim100$\AA~ in width that tends to fall in a random place along the spectrum depending on the position of the star within the field of view.

\begin{figure}
\epsscale{0.6}
\plotone{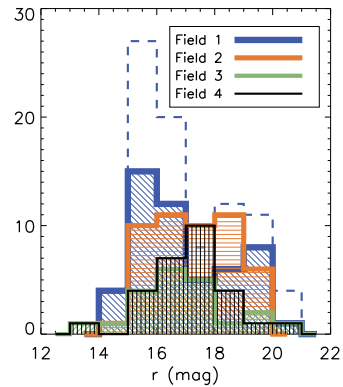}
\caption{The distribution of the $r$ magnitudes from BA14 for the detected sources with SNR$>$10 for each field. The dashed line shows the distribution of the total number of sources that were in the field 1.\label{fig:detected}}
\end{figure}

We observed 4 fields towards NGC 1980 using the multi-slit mode on January 7, 2016. We targeted 280 sources identified by BA14 with $r<$21 mag, of which 212 sources had probability of membership ($P$) greater than 50\%. The exposure time was 5$\times$10 min per field for the first three fields, and the last field had the exposure time of 8$\times$10 min. The weather conditions deteriorated rapidly over the course of the night, therefore, while the first field had the full sensitivity, the last field, even with the increased exposure time, yielded little in terms of the number of sources detected (Figure \ref{fig:detected}).

\begin{figure}
\epsscale{0.8}
\plotone{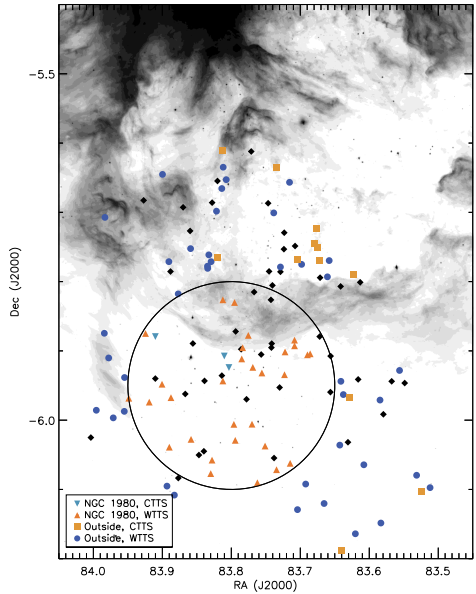}
\caption{Sources observed by this program. Sources that fall into the cirle are identified as members of the cluster (see Section \ref{sec:acc} for discussion). Black diamonds show the sources that are not part of the primary sample. The large circle is centered at $\alpha=83.8^\circ, \delta=-5.95^\circ$, center of NGC 1980 as defined by BA14, $0.15^\circ$ in radius. The greyscale background is 8 $\mu$m \textit{Spitzer} map from \citet{2012megeath}. \label{fig:map}}
\end{figure}

Partially because of this, we have detected only 148 sources (Figure \ref{fig:map}). However, the catalog of BA14 contains many spurious sources. Out of 165 sources that were targeted in the first two fields (of which 134 had $P>50$), only 103 (97 with $P>50$) have been detected. Similar trends persist in the other two fields within the sample limited to the detectable magnitudes. More than 20\% of sources with $P>50$, and the vast majority of sources with $P<50$ do not appear to be associated with physical objects, despite falling into the typically detectable magnitude range. As the catalog of BA14 was compiled from individual frames as opposed to the stacked images, many detections of cross-talks, saturation trails, as well as ghosts and cosmic rays may be included in the final catalog (Bouy, private communication).

\begin{figure}
\epsscale{0.8}
\plotone{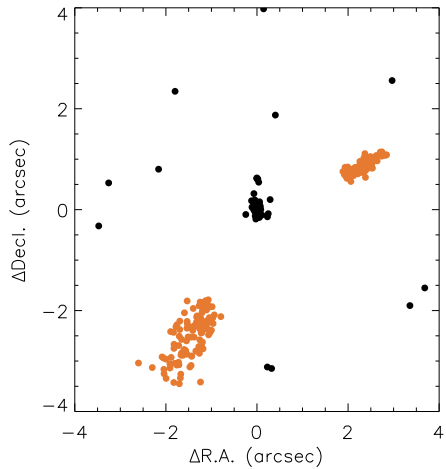}
\caption{Comparison of the astrometry between Gaia DR1 for the sources observed by this program, and all the sources that match its position in BA14. The sources concentrated near the center have accurate astrometry, and the two groups shown in orange consist mostly of duplicates.\label{fig:offset}}
\end{figure}

Several sources also appear to be duplicated in the catalog (some are listed as separate sources up to 3 times with the positional offset of up to 4", Figure \ref{fig:offset}). To correct the astrometry to remove this offset, we cross-matched the detected sources with the \textit{Gaia} DR1 source list \citep{gaiadr1}. All but three sources were matched with the DR1 list; for those that were not we used the astrometry from the initial Gaia source List \citep{2014smart}. We also re-cross-matched the sources with 2MASS photometry \citep{2mass}, as the positional offset has put many of the sources in BA14 outside of the cross-match radius, therefore, not all sources have their counterpart identified in that catalog. In addition, sources in BA14 that do have JHK photometry listed appear to have a systematic color offset from the 2MASS photometry.

\begin{figure}
	\gridline{\fig{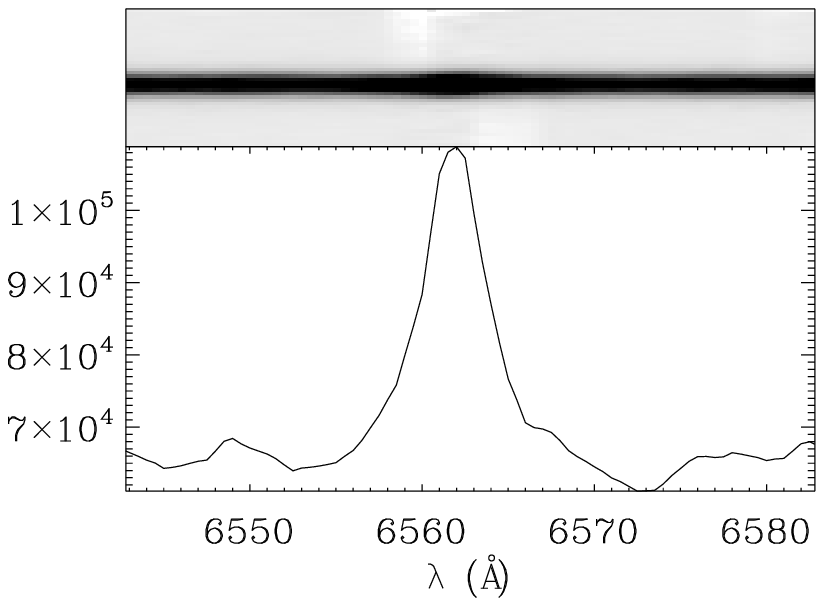}{0.16\textwidth}{}
			  \fig{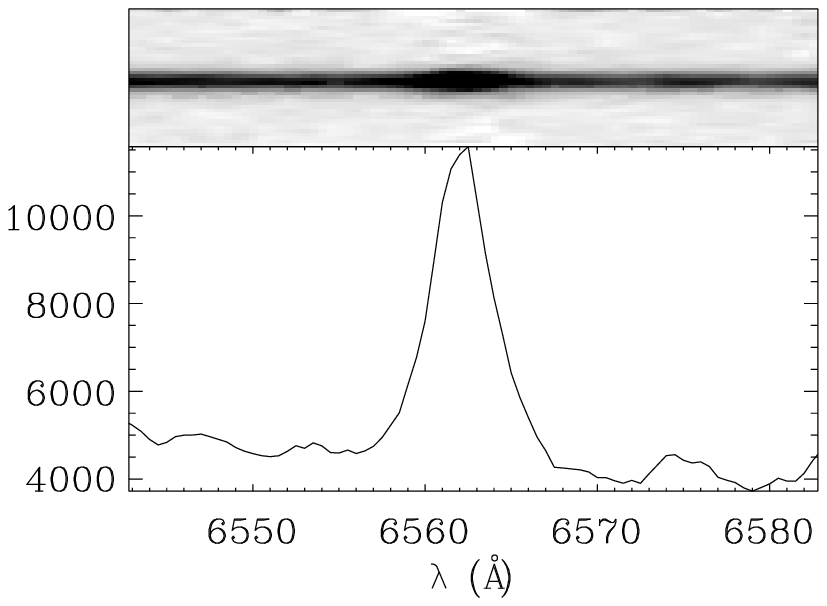}{0.16\textwidth}{}
			  \fig{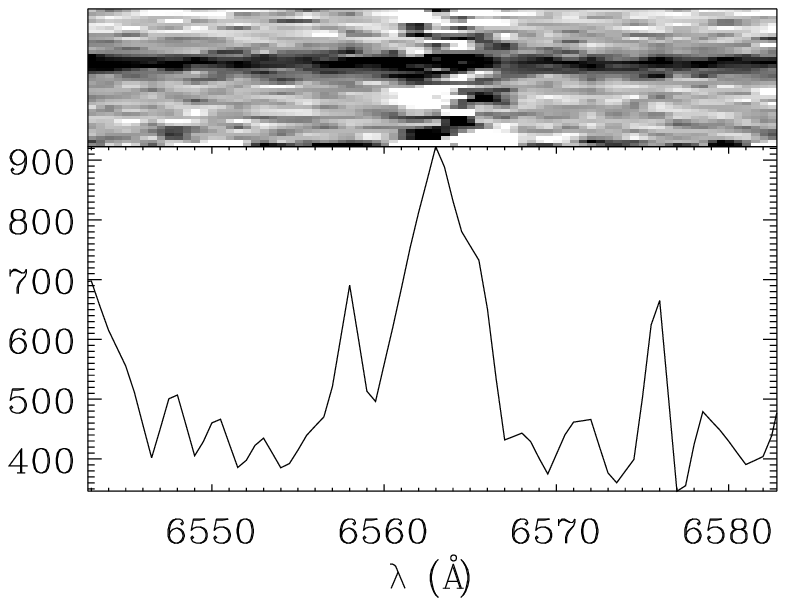}{0.15\textwidth}{}
             }
    \vspace{-0.8cm}
	\gridline{\hspace{-0.15cm}\fig{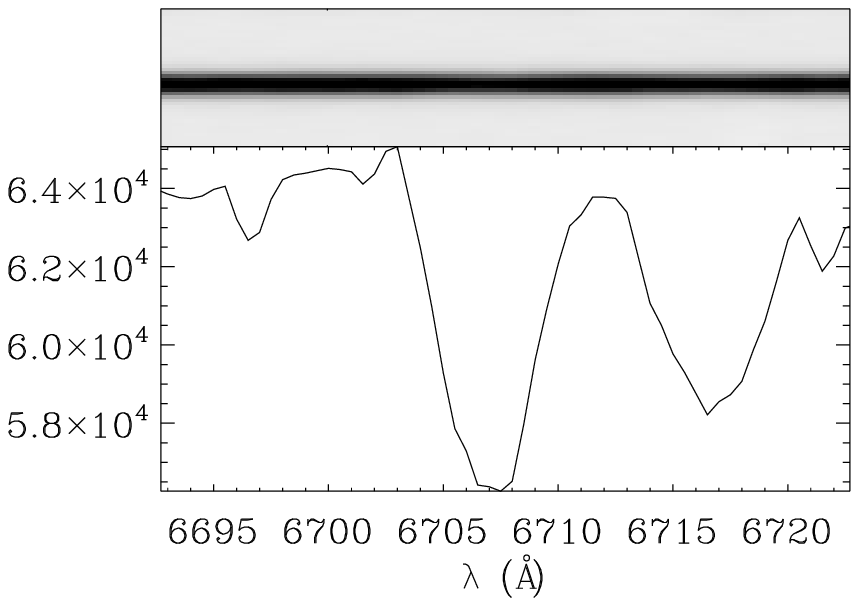}{0.17\textwidth}{}
			  \fig{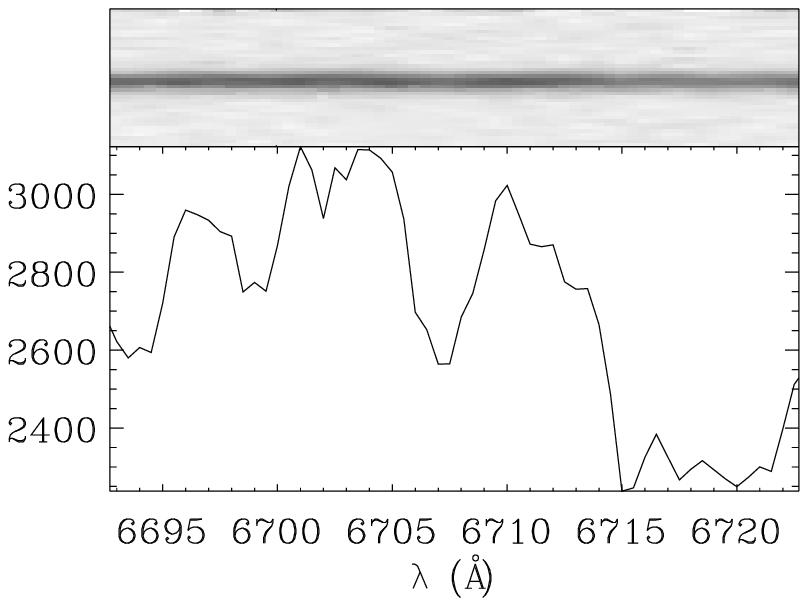}{0.16\textwidth}{}
			  \fig{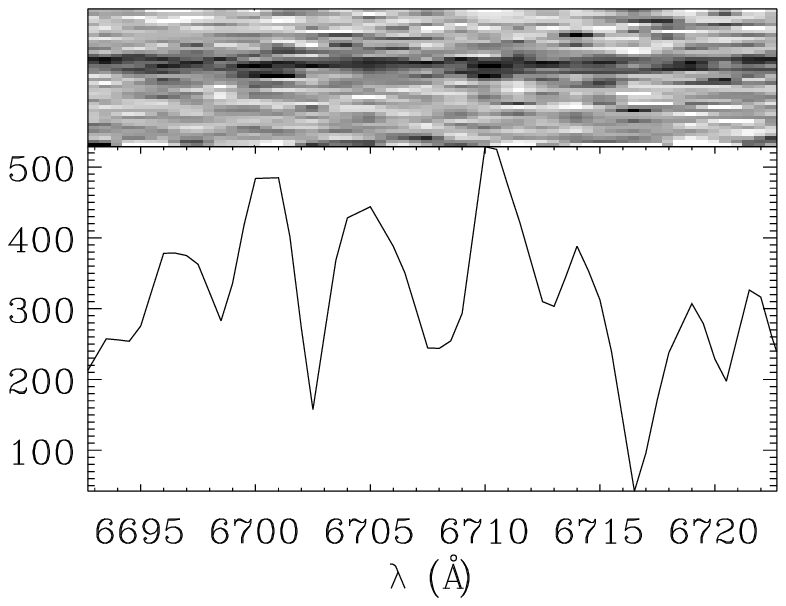}{0.155\textwidth}{}
             }
 \caption{Examples of high (BA 349989), medium (BA 346977), and low (373101) SNR spectra, showing H$\alpha$ and Li I lines. \label{fig:halicomp}}
\end{figure}

The IMACS spectra were reduced using the COSMOS pipeline\footnote{\url{http://code.obs.carnegiescience.edu/cosmos}} following the standard prescription. The final data product does have some issues with the poor subtraction of the nebular emission lines, particularly H$\alpha$ (6563\AA). In general, this emission originates from across the entire slit, and it is modeled and subtracted in a similar manner to the sky lines for the vast majority of sources. However, in some cases the nebular emission varies strongly across the slit, making subtraction difficult. This issue has a stronger effect on very low signal-to-noise (SNR) sources. We measured the equivalent width of H$\alpha$ ($W_{H_\alpha}$) in all spectra with an overall SNR$>5$. We visually examined the sky subtracted spectra and flagged the sources with poor subtraction, such as in cases where the flux of H$\alpha$ does not appear centered and/or confined to the star; in these cases we do not quote $W_{H_\alpha}$. To test nebular subtraction and to provide an estimate of the uncertainties of $W_{H_\alpha}$ we add up all flux along the slit outside of the source (normalizing it by the pixel count) to the flux of the source itself and measure $W_{H_\alpha}$ again. We do it in two different ways, by measuring a simple total of the flux as is, and by measuring a total of the absolute value of the flux, because due to variable nebular emission some of it may be oversubtracted. The uncertainties we report are the average difference between $W_{H_\alpha}$ we measure and $W_{H_\alpha}$ with the residuals added. Examples of the H$\alpha$ line profiles for low, medium, and high SNR data is shown in Figure \ref{fig:halicomp}.

The spectra were analyzed using SPTCLASS\footnote{\url{http://www.cida.gob.ve/~hernandj/SPTclass/sptclass.html}} \citep{2004Hernandez} in order to determine the spectral types of the detected stars. This program utilizes three different schemas optimized for three different ranges of stellar masses (K5 and later, late F to early K, and F5 and earlier) based on examining a number of spectral features. For late type stars (all the sources in this paper were identified as such) the examined features are TiO ($\lambda_{cen}$=4775, 4975, 5225, 5475, 5600, 5950, 6255, 6800, 7100, 7150\AA), and VO ($\lambda_{cen}$=7460, 7940, 7840, 8500, 8675, 8880\AA) lines. We report only on those spectral types that were obtained using at least three features.

\begin{figure}
\epsscale{1}
\plottwo{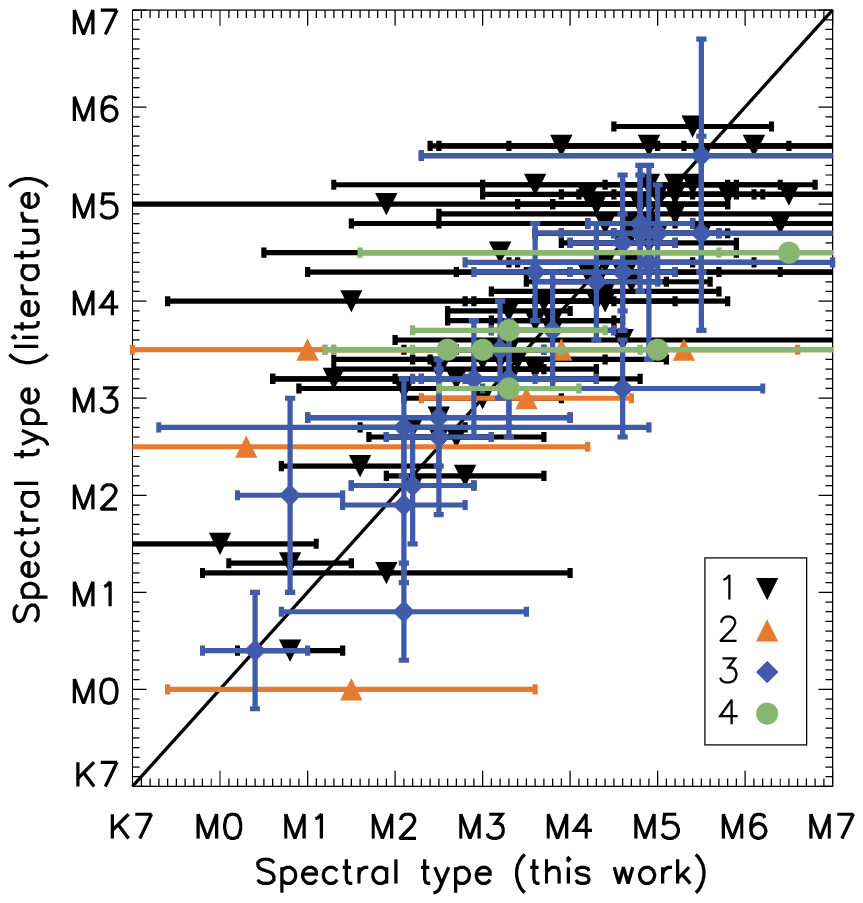}{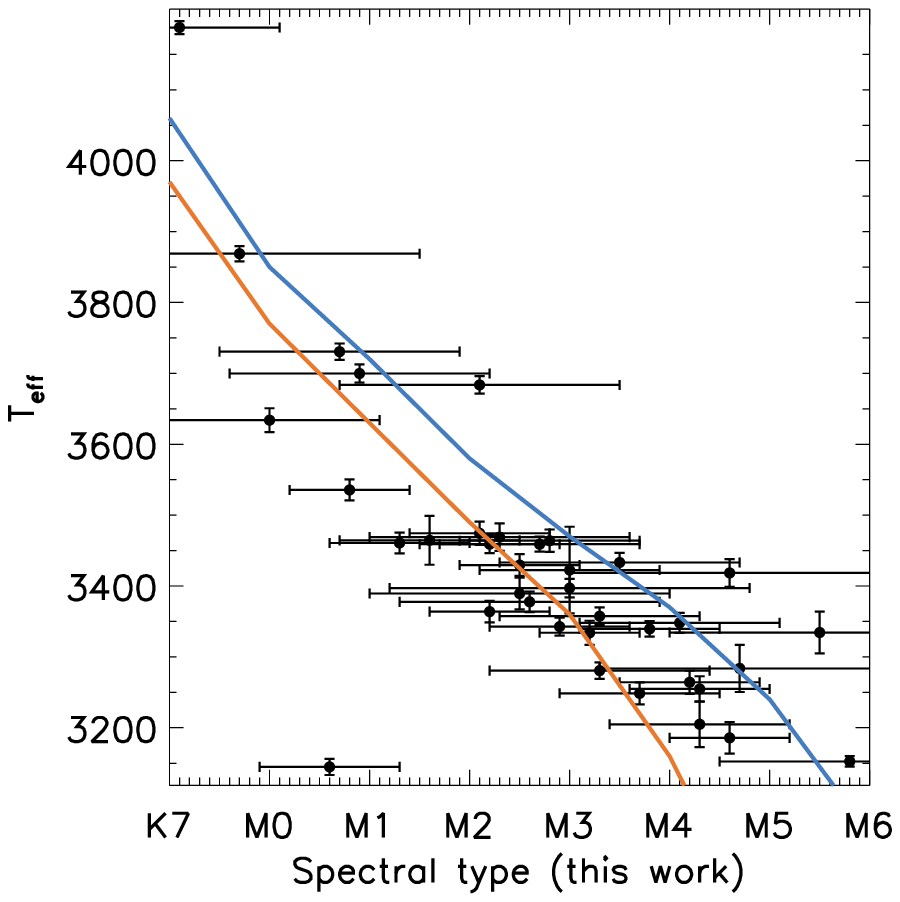}
\caption{Left panel: comparison of the measured spectral types to those that are available in literature. 1: \citet{2017fang}, 3: \citet{2000rebull} and \citet{2001rebull}, 3: \citet{2012Hsu}, 4: \citet{1997hillenbrand} and \citet{2010dario}. Right panel: comparison of the measured spectral types with the effective temperatures from \citet{2016dario}. Only sources with the spectral type measured better than two subclasses are included. The orange line shows spectral type to temperature conversion for young stars from \citet{2013Pecaut}, blue line from \citet{1995kenyon}. \label{fig:sptcomp}}
\end{figure}

Spectra for some stars have been analyzed in previous works. Spectral types for 25 stars that are in common with \citet{2012Hsu} and 76 stars in common with \citet{2017fang} show a good agreement within the uncertainties of each survey (usually within one subclass). Same can be said for 13 stars reported on by \citet{2010dario}, although due to their use of measurements compiled from various publications utilizing different methodologies, some additional scatter persists. Spectral types for 13 sources that are in common with \citet{2001rebull} show the largest scatter in the agreement; however, they are typically consistent with each other within the uncertainties (Figure \ref{fig:sptcomp}). Fifty stars presented in this work do not have pre-existing spectral classifications. We also compare the spectral types we measure to the effective temperature from \citet{2016dario}. Assuming typical spectral type to temperature conversions \citep[e.g][]{2013Pecaut}, they also show relatively good agreement with each other.

We also measured the equivalent widths of Li I line (6707\AA, $W_{Li}$), as it can be used as an indicator of youth and membership of star-forming regions \citep{1997briceno}. However, given the fact that it is a weak absorption line, due to the poor observing conditions in many sources the SNR was too small for a confident detection. In these cases the equivalent widths that can be measured from the random noise can be equivalent or greater than $W_{Li}$, which is typically expected to be $\sim$0.5\AA~ in the young stars. Therefore, we include the upper limits for sources where Li I line was not detected. We define the primary sample of the members consisting of sources with a confident Li I detection with $W_{Li}>0.1$ (62 stars), as well as the sources that have been previously identified as YSOs in other studies, have SNR$>5$, but it is not sufficiently high enough for a confident Li I detection (43 stars). The remaining 10 sources with SNR$>5$ are not included in the list of members; however, neither can they be conclusively rejected as such. Examples of Li I line profiles for low, medium, and high SNR data are shown in Figure \ref{fig:halicomp}.

\section{Discussion}
\subsection{Accretion}\label{sec:acc}

\begin{figure}
\epsscale{0.8}
\plotone{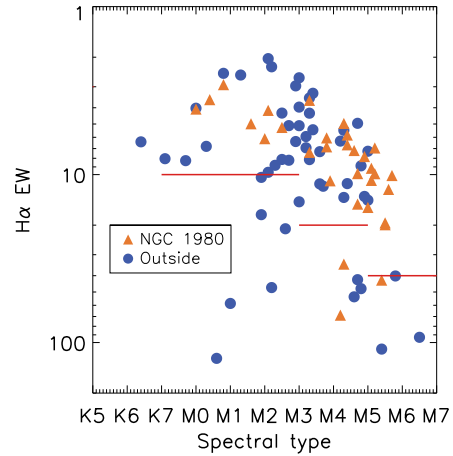}
\caption{H$\alpha$ equivalent width as a function of spectral type. Horizontal lines show the distinction between CTTS and WTTS objects based on the criteria by \citet{2003White}.\label{fig:ha}}
\end{figure}

We apply the criteria developed by \citet{2003White} to separate strongly accreting CTTSs from the WTTSs through analyzing $W_{H_\alpha}$ depending on the spectral type (Figure \ref{fig:ha}). Out of 85 stars in our total sample with measured $W_{H_\alpha}$, 15 are CTTS, for a CTTS fraction (defined as $N_{CTTS}/N_{CTTS+WTTS}$ of 0.18).

\citet{2013Pillitteri} have argued that NGC 1980 extends by $>1^\circ$ on the sky, but BA14 confine it to a much more limited area, possibly due to the difference in the methods, since BA14 rely on the background cloud to identify the sources. Nonetheless, we use their density map and define the sources within 0.15$^\circ$ from the center of the cluster ($\alpha=83.8^\circ, \delta=-5.95^\circ$ defined by BA14) as members, and the sources outside of that as those that are found on the periphery. There may be bona fide members of NGC 1980 outside of this boundary as well. Nonetheless, this boundary can be used to compare the population that exists at the center of the cluster to its surroundings. At the distance of 388 pc \citep{2017kounkel} the cluster diameter of 0.3$^\circ$ to a physical size of 2 pc. We find that within the cluster the CTTS fraction drops to only 0.09, where only three stars out of 33 show strong accretion signatures. On the other hand, outside of the cluster the CTTS fraction becomes 0.23 (12 out of 52 stars are accreting), which is dominated by the sources north of the cluster (0.27, 9 out of 33).

\citet{2017fang} argue that the median age of NGC 1980 is 1.8 Myr. Therefore the question arises, how significant is the difference that we observe in the CTTS fraction between the two populations. The sample in \citet{2017fang} is considerably larger, and it covers a bigger area of the ONC to the north, and L1641 to the south, in $-7^\circ<\delta<-4^\circ$ range, excluding the Trapezium cluster. The observed targets were also selected from BA14 catalog, however, there is not a direct 1--1 overlap between the surveys. In their sample there does appear to be little variation in the CTTS fraction of $\sim0.15$ throughout the observed regions, whether it is inside NGC 1980 or outside of it. A possible exception to this is NGC 1981 region, where the fraction does drop to $\sim$0.09. \citet{2017fang} do argue, that BA14 survey is biased against the sources with the hot inner disks; therefore the true CTTS fraction should be much higher than what can be observed.

To minimize this bias, we search for sources with H$\alpha$ observations in other surveys that fall into the NGC 1980 region. We include sources classified by \citet{2012Hsu}, \citet{2008furesz}, as well as analyze the H$\alpha$ measurements by \citet{2016kounkela}. In all the studies, we could identify 107 YSOs in the NGC 1980 region with H$\alpha$ line observed, out of which 31 are CTTS, for the total CTTS fraction of 0.29$\pm$0.05, assuming Gaussian statistics. 104 of these sources can be classified on the basis of the presence of a disk as indicated by infrared excesses detected with the IRAC camera on board the Spitzer Space Telescope \citep{2012megeath}. Out of 73 WTTSs, only 7 still have a disk, and out of 31 CTTSs, 7 sources do not have infrared excess which would be indicative of a full disk. The disk fraction for these sources is 0.30$\pm$0.05, very similar to what is found by \citet{2017fang} and \citet{2013Pillitteri}.

\subsection{Extinction}\label{sec:red}
\begin{figure}
	\gridline{\fig{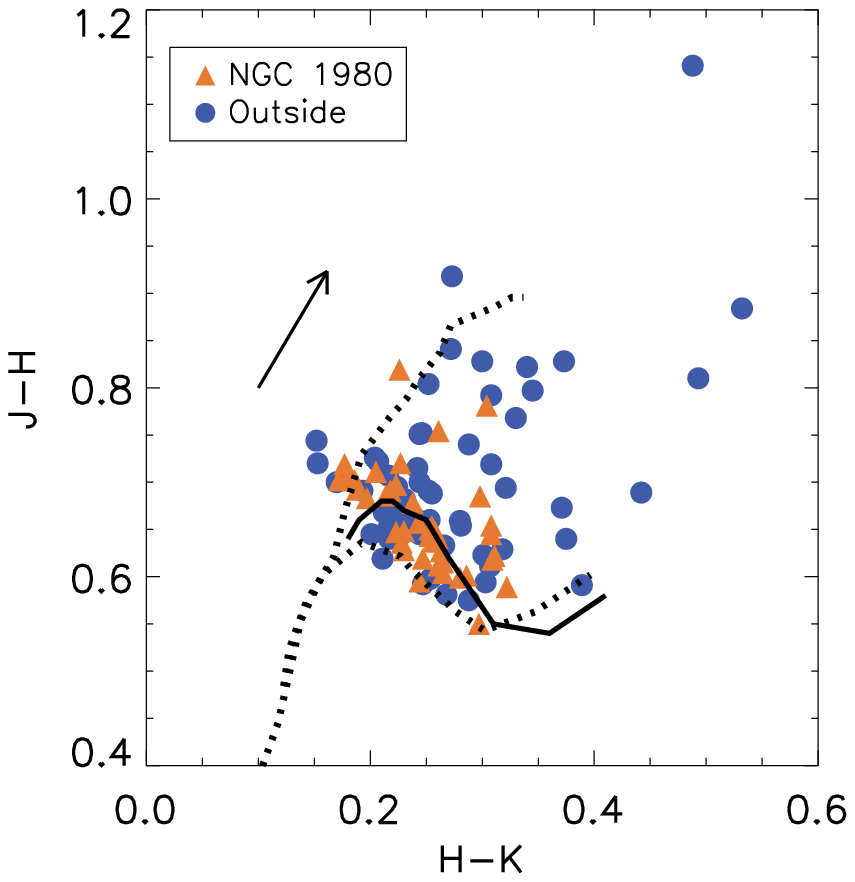}{0.25\textwidth}{}
			  \fig{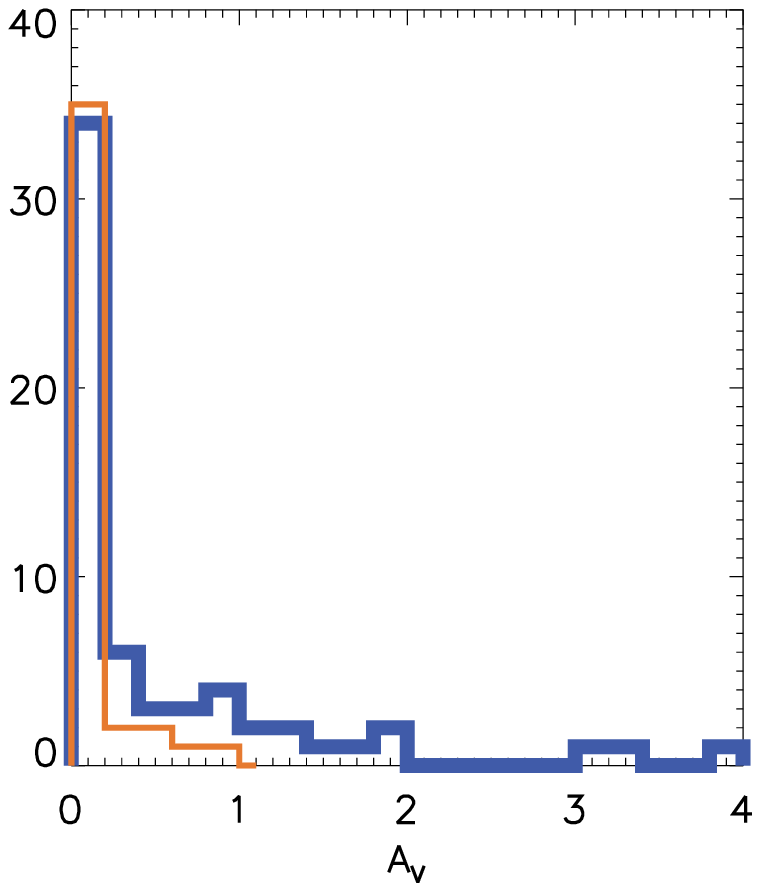}{0.22\textwidth}{}
             }
 \caption{Left: 2MASS J-H vs. H-K color-color diagram. Black dashed lines show typical colors for dwarfs (lower line) and giants (upper line) from \citet{1988bessell} with transformations from \citet{2001Carpenter}. Solid line showes the intrinsic colors of young stars from \citet{2013Pecaut}. Black arrow corresponds to the extinction of 1 A$_V$ \citep{2012megeath}. Left: distribution of A$_V$ measured towards the sources in the survey. Orange line corresponds to the sources within NGC 1980, blue to the sources located on the periphery. \label{fig:color}}
\end{figure}

We estimated the extinction towards the sources in the survey via their 2MASS photometry from comparing the J-H and H-K colors to the intrinsic colors of young (5-30 Myr) stars from \citet{2013Pecaut}. We adopt the relations of A$_J$:A$_K$=2.65, A$_J$:A$_K$=1.55 \citep{2012megeath}, and A$_K$:A$_V$=0.112 \citep{1989Cardelli}. Stars with very low extinctions may result in negative A$_V$ measurements, in those cases A$_V$ is set to 0. We report on the average A$_V$ measured from both J-H and H-K colors for each star in Table \ref{tab:main} (Figure \ref{fig:color}). The typical error due to an uncertainty in spectral type is A$_V\sim$0.1. We have also attempted to measure A$_V$ through artificially de-redenning the spectra \citep[assuming the standard extinction law $R_V$=3.1,][]{1989Cardelli} until the slope matches that of the template spectra of the corresponding spectral type from the X-Shooter Spectral Library \citep{x-shooter}. While this method does produce mostly comparable A$_V$ measurements, they are also significantly more uncertain.

In general approximately half of all sources do have very low extinction with A$_V\sim0$. This is true of almost all sources identified as members of NGC 1980, with only six sources (out of 39) with A$_V>0.1$, and all with A$_V<1.0$.  The degree of extinction towards the cluster comparable to what has been found by \citet{2013Pillitteri} through X-ray observations of the region. On the periphery of the cluster, 32 out of 58 sources have A$_V>0.1$, 11 of which have A$_V>1.0$. The Kolmogorov-Smirnov test shows that the two populations are different by 3.5$\sigma$.

Most of the sources with very high extinction in our survey originate north of the NGC 1980 cluster, towards the ONC. The A$_V$ distribution of these northernmost sources does not mirror that measured towards the ONC by \citet{2010dario}, however, the sample selection of the two surveys is very different and our optical study is biased against highly extincted sources. However, in general our results are consistent with what has been observed in the other works. The extinction in the Orion A molecular cloud is low in the vicinity of the NGC 1980, it rises sharply towards the ONC, and gradually increases southward along the L1641 filament \citep{2010dario,2012Hsu,2013Pillitteri,2017fang}.

\subsection{Li I distribution}\label{sec:li}

\begin{figure}
\epsscale{0.8}
\plotone{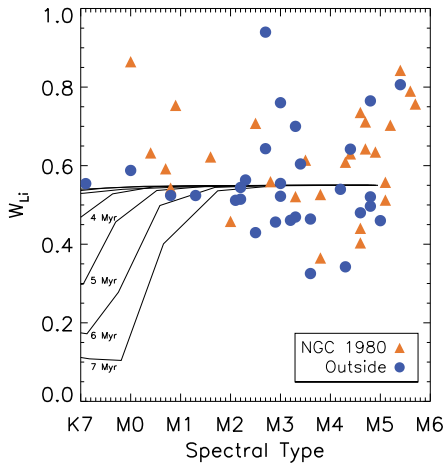}
\caption{Distribution of $W_{Li}$ as a function of spectral type. Black lines show the evolution of $W_{Li}$ from \citet{2015baraffe} tracks, from 1 to 7 Myr, converted using transformations from \citet{2013Pecaut}. The typical uncertainty in measured $W_{Li}$ is $\sim$0.1\AA.\label{fig:li}}
\end{figure}

As young stars evolve, eventually they will begin depleting Li, and more massive stars will deplete it faster. Therefore, in some cases the distribution of $W_{Li}$ as a function of temperature can be used as an additional information from which age can be inferred \citep[e.g.][]{2017Jeffries}. We compare the measured $W_{Li}$ for the sources with firm Li I detections to the expected Li I distribution from the evolutionary tracks from \citet{2015baraffe}. Nearly all of our sample consists of M type stars; these sources do not begin processing Li I until 5--10 Myr. From this comparison, we can rule out ages older than about 6 Myr for the sources found towards NGC 1980, however, this comparison cannot help to distinguish ages younger than that (Figure \ref{fig:li}). A similar distribution of $W_{Li}$ is found in the surveys by \citet{2012Hsu} and \citet{2017fang} throughout the ONC and L1641.

\section{Conclusion}

We obtained low-resolution spectra from IMACS towards 148 sources towards NGC 1980, we determined spectral types of these sources, and determined whether or not they can be considered as YSOs on the basis of Li I absorption. Most sources have A$_V\sim0$, in agreement with what have been found by \citet{2013Pillitteri} and BA14. We determined whether or not these sources have H$\alpha$ excess due to the ongoing accretion, and searched for sources that fall into the region identified as NGC 1980 with H$\alpha$ measurements in the previous works. We found that 31 out of 107 stars in the cluster can be identified as accretors, for a CTTS fraction of 0.29$\pm$0.05.

As the population ages, the CTTS fraction decreases \citep[e.g.][Brice\~{n}o et al. in prep]{2009mamajek}. Within the ONC, containing large population that spans in age between $<1-3$ Myr, the CTTS fraction is $\sim$0.5 \citep{2008furesz,2009dario}. Within the outlying L1641 cloud it decreases to is 0.35 ($\sim$3 Myr) \citep{2012Hsu}. Within the Ori OB1b sub-association (4 Myr), CTTS fraction is 0.13, and only 0.06\% towards 25 Ori (8 Myr) comoving group. \citep{2007Briceno}.

From this, it does appear that the population of NGC 1980 is indistinguishable from the population of L1641. This is similar to what has been observed by \citet{2017fang}. However, the CTTS fraction is considerably lower in this cluster than in the ONC itself. It is possible that the sample is highly incomplete. Given that all classifications were conducted from an optical surveys, there exists an inherent bias against deeply embedded and accreting sources. On the other hand, WTTSs typically have a much smaller IR excess, making them harder to identify as targets when planning a survey. Since the extinction in the vicinity of NGC 1980 is much lower than in the ONC, it would be unlikely for there to be a larger population of highly reddened undetected sources to be present in NGC 1980 over the ONC. Unless the CTTS fraction is significantly lower in the ONC due to a presence of previously unanalyzed WTTSs, NGC 1980 is probably somewhat older. It is unlikely to be as old as 5 Myr (the age that was previously estimated by BA14). We estimate its age to be $\sim$3 Myr, similar to L1641. A possible reason why \citet{2017fang} do not see the difference between the ONC and NGC 1980 is because the sample for which they determined the ages is highly incomplete near the ONC (Trapezium and OMC 2/3) region, the youngest and most deeply embedded population. And, with the updated distances from \citet{2017kounkel}, the ages measured by \citet{2017fang} would be $\sim$20\% older, which would also help in reconciling some of the differences. 

As this work does not explicitly measure distances to the stellar populations, in this work we cannot conclusively confirm or disprove whether NGC 1980 is in the foreground of the ONC or not. The reason for low extinction near the region could be that cluster is located on the outer layer of the molecular cloud, in which case it would technically not be incorrect to refer to the cluster as a foreground population by the slimmest of margins. However, a more likely explanation would be that the stellar winds have cleared out the molecular gas, there could have been less gas near the cluster than there was near the ONC from the very beginning. \citet{2017kounkel} have shown that the ONC is found at the same distance that was suggested by BA14 for NGC 1980. \citet{2017kounkel} did not detect any of the stars towards NGC 1980 in their sample, therefore they were not able to estimate stellar parallax towards the region. The distance found by BA14 was found photometrically, therefore it has significantly larger uncertainty in its value.

Until the uncertainties in the distance are smaller than any distance offsets, it is not entirely meaningful to discuss substructure, as well as what is and what is not in the foreground. The future data releases from \textit{Gaia} space telescope should would be instrumental in measuring the three-dimensional relations and intercluster kinematics in the region, and it should resolve this matter once and for all. Its observations should shed light on the initial conditions that have led to the current configurations of these star forming regions.

\acknowledgments
The authors would like to thank the LCO operators and staff for their help on the observing run. MK acknowledges the insightful conversation with Min Fang. We also acknowledge a prompt and helpful report from an anonymous referee. This research made use of the cross-match service provided by CDS, Strasbourg. This work was supported in part by the University of Michigan.

\software{TOPCAT \citep{topcat}, COSMOS}
\facility{Magellan: Baade(IMACS)}

\bibliographystyle{apj.bst}

\begin{thebibliography}{}
\expandafter\ifx\csname natexlab\endcsname\relax\def\natexlab#1{#1}\fi

\bibitem[{{Alves} \& {Bouy}(2012)}]{2012alves}
{Alves}, J., \& {Bouy}, H. 2012, \aap, 547, A97

\bibitem[{{Baraffe} {et~al.}(2015){Baraffe}, {Homeier}, {Allard}, \&
  {Chabrier}}]{2015baraffe}
{Baraffe}, I., {Homeier}, D., {Allard}, F., \& {Chabrier}, G. 2015, \aap, 577,
  A42

\bibitem[{{Bessell} \& {Brett}(1988)}]{1988bessell}
{Bessell}, M.~S., \& {Brett}, J.~M. 1988, \pasp, 100, 1134

\bibitem[{{Bigelow} \& {Dressler}(2003)}]{imacs}
{Bigelow}, B.~C., \& {Dressler}, A.~M. 2003, in \procspie, Vol. 4841,
  Instrument Design and Performance for Optical/Infrared Ground-based
  Telescopes, ed. M.~{Iye} \& A.~F.~M. {Moorwood}, 1727--1738

\bibitem[{{Bouy} {et~al.}(2014){Bouy}, {Alves}, {Bertin}, {Sarro}, \&
  {Barrado}}]{2014bouy}
{Bouy}, H., {Alves}, J., {Bertin}, E., {Sarro}, L.~M., \& {Barrado}, D. 2014,
  \aap, 564, A29

\bibitem[{{Brice{\~n}o} {et~al.}(2007){Brice{\~n}o}, {Hartmann},
  {Hern{\'a}ndez}, {Calvet}, {Vivas}, {Furesz}, \&
  {Szentgyorgyi}}]{2007Briceno}
{Brice{\~n}o}, C., {Hartmann}, L., {Hern{\'a}ndez}, J., {et~al.} 2007, \apj,
  661, 1119

\bibitem[{{Briceno} {et~al.}(1997){Briceno}, {Hartmann}, {Stauffer}, {Gagne},
  {Stern}, \& {Caillault}}]{1997briceno}
{Briceno}, C., {Hartmann}, L.~W., {Stauffer}, J.~R., {et~al.} 1997, \aj, 113,
  740

\bibitem[{{Cardelli} {et~al.}(1989){Cardelli}, {Clayton}, \&
  {Mathis}}]{1989Cardelli}
{Cardelli}, J.~A., {Clayton}, G.~C., \& {Mathis}, J.~S. 1989, \apj, 345, 245

\bibitem[{{Carpenter}(2001)}]{2001Carpenter}
{Carpenter}, J.~M. 2001, \aj, 121, 2851

\bibitem[{{Chen} {et~al.}(2014){Chen}, {Trager}, {Peletier}, {Lan{\c c}on},
  {Vazdekis}, {Prugniel}, {Silva}, \& {Gonneau}}]{x-shooter}
{Chen}, Y.-P., {Trager}, S.~C., {Peletier}, R.~F., {et~al.} 2014, \aap, 565,
  A117

\bibitem[{{Cutri} {et~al.}(2003){Cutri}, {Skrutskie}, {van Dyk}, {Beichman},
  {Carpenter}, {Chester}, {Cambresy}, {Evans}, {Fowler}, {Gizis}, {Howard},
  {Huchra}, {Jarrett}, {Kopan}, {Kirkpatrick}, {Light}, {Marsh}, {McCallon},
  {Schneider}, {Stiening}, {Sykes}, {Weinberg}, {Wheaton}, {Wheelock}, \&
  {Zacarias}}]{2mass}
{Cutri}, R.~M., {Skrutskie}, M.~F., {van Dyk}, S., {et~al.} 2003, VizieR Online
  Data Catalog, 2246, 0

\bibitem[{{Da Rio} {et~al.}(2009){Da Rio}, {Robberto}, {Soderblom}, {Panagia},
  {Hillenbrand}, {Palla}, \& {Stassun}}]{2009dario}
{Da Rio}, N., {Robberto}, M., {Soderblom}, D.~R., {et~al.} 2009, \apjs, 183,
  261

\bibitem[{{Da Rio} {et~al.}(2010){Da Rio}, {Robberto}, {Soderblom}, {Panagia},
  {Hillenbrand}, {Palla}, \& {Stassun}}]{2010dario}
---. 2010, \apj, 722, 1092

\bibitem[{{Da Rio} {et~al.}(2016){Da Rio}, {Tan}, {Covey}, {Cottaar}, {Foster},
  {Cullen}, {Tobin}, {Kim}, {Meyer}, {Nidever}, {Stassun}, {Chojnowski},
  {Flaherty}, {Majewski}, {Skrutskie}, {Zasowski}, \& {Pan}}]{2016dario}
{Da Rio}, N., {Tan}, J.~C., {Covey}, K.~R., {et~al.} 2016, \apj, 818, 59

\bibitem[{{de Zeeuw} {et~al.}(2001){de Zeeuw}, {Hoogerwerf}, \& {de
  Bruijne}}]{2001deZeeuw}
{de Zeeuw}, T., {Hoogerwerf}, R., \& {de Bruijne}, J. 2001, in Astronomical
  Society of the Pacific Conference Series, Vol. 228, Dynamics of Star Clusters
  and the Milky Way, ed. S.~{Deiters}, B.~{Fuchs}, A.~{Just}, R.~{Spurzem}, \&
  R.~{Wielen}, 201

\bibitem[{{Fang} {et~al.}(2009){Fang}, {van Boekel}, {Wang}, {Carmona},
  {Sicilia-Aguilar}, \& {Henning}}]{2009Fang}
{Fang}, M., {van Boekel}, R., {Wang}, W., {et~al.} 2009, \aap, 504, 461

\bibitem[{{Fang} {et~al.}(2017){Fang}, {Kim}, {Pascucci}, {Apai}, {Zhang},
  {Sicilia-Aguilar}, {Alonso-Mart{\'{\i}}nez}, {Eiroa}, \& {Wang}}]{2017fang}
{Fang}, M., {Kim}, J.~S., {Pascucci}, I., {et~al.} 2017, ArXiv e-prints,
  arXiv:1703.06948

\bibitem[{{F{\H u}r{\'e}sz} {et~al.}(2008){F{\H u}r{\'e}sz}, {Hartmann},
  {Megeath}, {Szentgyorgyi}, \& {Hamden}}]{2008furesz}
{F{\H u}r{\'e}sz}, G., {Hartmann}, L.~W., {Megeath}, S.~T., {Szentgyorgyi},
  A.~H., \& {Hamden}, E.~T. 2008, \apj, 676, 1109

\bibitem[{{Gaia Collaboration} {et~al.}(2016){Gaia Collaboration}, {Brown},
  {Vallenari}, {Prusti}, {de Bruijne}, {Mignard}, {Drimmel}, {Babusiaux},
  {Bailer-Jones}, {Bastian}, \& et~al.}]{gaiadr1}
{Gaia Collaboration}, {Brown}, A.~G.~A., {Vallenari}, A., {et~al.} 2016, \aap,
  595, A2

\bibitem[{{Gualandris} {et~al.}(2004){Gualandris}, {Portegies Zwart}, \&
  {Eggleton}}]{2004Gualandris}
{Gualandris}, A., {Portegies Zwart}, S., \& {Eggleton}, P.~P. 2004, \mnras,
  350, 615

\bibitem[{{Hern{\'a}ndez} {et~al.}(2004){Hern{\'a}ndez}, {Calvet},
  {Brice{\~n}o}, {Hartmann}, \& {Berlind}}]{2004Hernandez}
{Hern{\'a}ndez}, J., {Calvet}, N., {Brice{\~n}o}, C., {Hartmann}, L., \&
  {Berlind}, P. 2004, \aj, 127, 1682

\bibitem[{{Hillenbrand}(1997)}]{1997hillenbrand}
{Hillenbrand}, L.~A. 1997, \aj, 113, 1733

\bibitem[{{Hsu} {et~al.}(2012){Hsu}, {Hartmann}, {Allen}, {Hern{\'a}ndez},
  {Megeath}, {Mosby}, {Tobin}, \& {Espaillat}}]{2012Hsu}
{Hsu}, W.-H., {Hartmann}, L., {Allen}, L., {et~al.} 2012, \apj, 752, 59

\bibitem[{{Hsu} {et~al.}(2013){Hsu}, {Hartmann}, {Allen}, {Hern{\'a}ndez},
  {Megeath}, {Tobin}, \& {Ingleby}}]{2013Hsu}
---. 2013, \apj, 764, 114

\bibitem[{{Jeffries} {et~al.}(2017){Jeffries}, {Jackson}, {Franciosini},
  {Randich}, {Barrado}, {Frasca}, {Klutsch}, {Lanzafame}, {Prisinzano},
  {Sacco}, {Gilmore}, {Vallenari}, {Alfaro}, {Koposov}, {Pancino}, {Bayo},
  {Casey}, {Costado}, {Damiani}, {Hourihane}, {Lewis}, {Jofre}, {Magrini},
  {Monaco}, {Morbidelli}, {Worley}, {Zaggia}, \& {Zwitter}}]{2017Jeffries}
{Jeffries}, R.~D., {Jackson}, R.~J., {Franciosini}, E., {et~al.} 2017, \mnras,
  464, 1456

\bibitem[{{Kenyon} \& {Hartmann}(1995)}]{1995kenyon}
{Kenyon}, S.~J., \& {Hartmann}, L. 1995, \apjs, 101, 117

\bibitem[{{Kounkel} {et~al.}(2016){Kounkel}, {Hartmann}, {Tobin}, {Mateo},
  {Bailey}, \& {Spencer}}]{2016kounkela}
{Kounkel}, M., {Hartmann}, L., {Tobin}, J.~J., {et~al.} 2016, \apj, 821, 8

\bibitem[{{Kounkel} {et~al.}(2017){Kounkel}, {Hartmann}, {Loinard},
  {Ortiz-Le{\'o}n}, {Mioduszewski}, {Rodr{\'{\i}}guez}, {Dzib}, {Torres},
  {Pech}, {Galli}, {Rivera}, {Boden}, {Evans}, {Brice{\~n}o}, \&
  {Tobin}}]{2017kounkel}
{Kounkel}, M., {Hartmann}, L., {Loinard}, L., {et~al.} 2017, \apj, 834, 142

\bibitem[{{Mamajek}(2009)}]{2009mamajek}
{Mamajek}, E.~E. 2009, in American Institute of Physics Conference Series, Vol.
  1158, American Institute of Physics Conference Series, ed. T.~{Usuda},
  M.~{Tamura}, \& M.~{Ishii}, 3--10

\bibitem[{{Megeath} {et~al.}(2012){Megeath}, {Gutermuth}, {Muzerolle},
  {Kryukova}, {Flaherty}, {Hora}, {Allen}, {Hartmann}, {Myers}, {Pipher},
  {Stauffer}, {Young}, \& {Fazio}}]{2012megeath}
{Megeath}, S.~T., {Gutermuth}, R., {Muzerolle}, J., {et~al.} 2012, \aj, 144,
  192

\bibitem[{{Menten} {et~al.}(2007){Menten}, {Reid}, {Forbrich}, \&
  {Brunthaler}}]{2007menten}
{Menten}, K.~M., {Reid}, M.~J., {Forbrich}, J., \& {Brunthaler}, A. 2007, \aap,
  474, 515

\bibitem[{{Morales-Calder{\'o}n} {et~al.}(2012){Morales-Calder{\'o}n},
  {Stauffer}, {Stassun}, {Vrba}, {Prato}, {Hillenbrand}, {Terebey}, {Covey},
  {Rebull}, {Terndrup}, {Gutermuth}, {Song}, {Plavchan}, {Carpenter},
  {Marchis}, {Garc{\'{\i}}a}, {Margheim}, {Luhman}, {Angione}, \&
  {Irwin}}]{2012Morales}
{Morales-Calder{\'o}n}, M., {Stauffer}, J.~R., {Stassun}, K.~G., {et~al.} 2012,
  \apj, 753, 149

\bibitem[{{Pecaut} \& {Mamajek}(2013)}]{2013Pecaut}
{Pecaut}, M.~J., \& {Mamajek}, E.~E. 2013, \apjs, 208, 9

\bibitem[{{Pillitteri} {et~al.}(2013){Pillitteri}, {Wolk}, {Megeath}, {Allen},
  {Bally}, {Gagn{\'e}}, {Gutermuth}, {Hartman}, {Micela}, {Myers}, {Oliveira},
  {Sciortino}, {Walter}, {Rebull}, \& {Stauffer}}]{2013Pillitteri}
{Pillitteri}, I., {Wolk}, S.~J., {Megeath}, S.~T., {et~al.} 2013, \apj, 768, 99

\bibitem[{{Rebull}(2001)}]{2001rebull}
{Rebull}, L.~M. 2001, \aj, 121, 1676

\bibitem[{{Rebull} {et~al.}(2000){Rebull}, {Hillenbrand}, {Strom}, {Duncan},
  {Patten}, {Pavlovsky}, {Makidon}, \& {Adams}}]{2000rebull}
{Rebull}, L.~M., {Hillenbrand}, L.~A., {Strom}, S.~E., {et~al.} 2000, \aj, 119,
  3026

\bibitem[{{Smart} \& {Nicastro}(2014)}]{2014smart}
{Smart}, R.~L., \& {Nicastro}, L. 2014, \aap, 570, A87

\bibitem[{{Taylor}(2005)}]{topcat}
{Taylor}, M.~B. 2005, in Astronomical Society of the Pacific Conference Series,
  Vol. 347, Astronomical Data Analysis Software and Systems XIV, ed.
  P.~{Shopbell}, M.~{Britton}, \& R.~{Ebert}, 29

\bibitem[{{White} \& {Basri}(2003)}]{2003White}
{White}, R.~J., \& {Basri}, G. 2003, \apj, 582, 1109

\end{thebibliography}

\begin{deluxetable*}{cccccccccccc}
\tabletypesize{\scriptsize}
\tablewidth{0pt}
\tablecaption{Sources observed towards NGC 1980 with IMACS. \label{tab:main}}
\tablehead{
\colhead{BA14} &\colhead{R.A.} &\colhead{Decl.} & \colhead{SPType} & \colhead{$W_{H_\alpha}$\tablenotemark{a}} & \colhead{$W_{Li}$\tablenotemark{a}}& \colhead{SNR} & \colhead{Accr.} & \colhead{Reference\tablenotemark{b}} & \colhead{A$_V$} & \colhead{Field}\\
\colhead{$\#$} &\colhead{J2000.0} &\colhead{J2000.0} & \colhead{} & \colhead{\AA} & \colhead{\AA}& \colhead{} & \colhead{} & \colhead{} & \colhead{mag.} & \colhead{}
}
\startdata
329411 & 05:34:02.85 & -06:05:50.7 & M4.8$\pm$0.9 &   -8.9$\pm$ 0.2 &     0.5 &  59.4 & WTTS & 1~3~~~7 & 0.2 & 1 \\
331374 & 05:34:05.99 & -06:06:11.7 & M4.6$\pm$0.6 &  -53.4$\pm$ 2.7 &     0.5 &  66.2 & CTTS & 1~3~5~7 & 0.0 & 1 \\
332392 & 05:34:07.52 & -06:04:46.4 & M2.5$\pm$0.6 &   -4.3$\pm$ 0.2 &     0.4 & 170.5 & WTTS & 1~3~~~7 & 0.0 & 1 \\
340962 & 05:34:19.95 & -06:08:54.6 & M3.3$\pm$0.5 &   -3.5$\pm$ 0.2 &     0.5 & 116.9 & WTTS & 1~3~~~~ & 0.0 & 1 \\
341157 & 05:34:20.26 & -05:58:15.6 & M4.7$\pm$1.4 &   -5.0$\pm$ 1.2 & $<$ 0.2 &  29.0 & WTTS & ~~~4~~7 & 0.0 & 1 \\
344878 & 05:34:25.89 & -06:03:51.9 & M5.0$\pm$0.5 &  -14.2$\pm$ 0.6 &        \tablenotemark{c} &  75.0 & WTTS & ~~3~~~7 & 0.1 & 1 \\
346977 & 05:34:28.84 & -06:09:50.4 & M5.0$\pm$0.7 &   -7.3$\pm$ 0.4 &     0.5 &  35.3 & WTTS & 1~~~~~~ & 0.8 & 1 \\
348520 & 05:34:30.97 & -05:58:03.6 & M6.5$\pm$1.3 &  -93.1$\pm$33.2 & $<$ 1.3 &   5.6 & CTTS & ~~~~56~ & 0.0 & 1 \\
348805 & 05:34:31.41 & -06:01:54.7 & M7.5$\pm$5.1 &    --- &     --- &   4.0 & --- & ~~~~~~~ & --- & 1 \\
349989 & 05:34:33.02 & -05:57:47.0 & M0.0$\pm$0.6 &   -4.0$\pm$ 1.1 &     0.6 & 172.8 & WTTS & 12~4~6~ & 1.5 & 1 \\
350399 & 05:34:33.61 & -06:11:19.4 & M4.8$\pm$0.6 &  -47.8$\pm$ 2.9 &     0.5 &  59.0 & CTTS & 1~3~5~7 & 0.2 & 1 \\
350559 & 05:34:33.87 & -05:56:38.1 & M4.2$\pm$0.7 &   -6.3$\pm$ 1.2 &     0.5 & 109.8 & WTTS & 12~~~67 & 0.0 & 1 \\
354860 & 05:34:39.72 & -06:07:13.2 & M3.2$\pm$0.5 &   -5.9$\pm$ 0.4 &     0.5 & 154.7 & WTTS & 1~34~~7 & 0.0 & 1 \\
354978 & 05:34:39.87 & -06:08:34.1 & M2.1$\pm$2.8 &    \tablenotemark{c} &        \tablenotemark{c} &  52.9 & --- & ~23~~~~ & 0.0 & 1 \\
358064 & 05:34:44.56 & -05:54:15.2 & M5.7$\pm$0.8 &  -10.1$\pm$ 1.1 &     0.8 &  12.1 & WTTS & 1~~~~~~ & 0.0 & 1 \\
358673 & 05:34:45.51 & -05:54:20.5 & M4.4$\pm$0.8 &   -5.8$\pm$ 1.3 &     0.6 &  60.4 & WTTS & 12~~~~7 & 0.0 & 1 \\
359072 & 05:34:46.13 & -06:05:32.0 & M2.1$\pm$0.7 &   -2.0$\pm$ 0.7 &     0.5 & 147.4 & WTTS & 1~~~~~~ & 0.0 & 1 \\
361022 & 05:34:48.98 & -06:07:45.0 & M2.2$\pm$0.7 &   -2.3$\pm$ 0.7 &     0.5 & 152.0 & WTTS & 1~3~~~7 & 0.0 & 1 \\
361725 & 05:34:50.01 & -05:53:33.6 & M5.6$\pm$0.7 &  -12.3$\pm$ 1.0 &     0.8 &  16.9 & WTTS & 1~~~~~~ & 0.0 & 1 \\
362799 & 05:34:51.53 & -06:03:44.4 & M3.5$\pm$4.4 &    4.9$\pm$ 1.1 &     0.6 &  17.2 & WTTS & 1~~4~~~ & 0.9 & 1 \\
363805 & 05:34:53.31 & -05:54:05.4 & M3.8$\pm$0.7 &   -6.8$\pm$ 1.1 &     0.4 &  95.5 & WTTS & 12~4~67 & 0.5 & 1 \\
363908 & 05:34:53.49 & -05:56:03.7 & M5.1$\pm$0.8 &  -10.8$\pm$ 1.0 &     0.6 &  21.7 & WTTS & 1~~~~~~ & 0.0 & 1 \\
365030 & 05:34:55.20 & -05:57:10.1 & M5.8$\pm$1.0 &  -10.6$\pm$ 1.5 & $<$ 0.3 &  14.1 & --- & ~~~~~~~ & 0.0 & 1 \\
365208 & 05:34:55.64 & -06:01:03.7 & M4.6$\pm$1.6 &    --- &     0.7 &  13.0 & --- & 123~~~~ & 0.0 & 1 \\
365598 & 05:34:56.27 & -06:04:17.4 & M2.1$\pm$1.4 &   -4.1$\pm$ 0.7 & $<$ 0.2 &  19.7 & WTTS & ~23~~~~ & 0.0 & 1 \\
366171 & 05:34:57.13 & -06:03:16.9 & M6.5$\pm$1.1 &  -38.1$\pm$12.2 & $<$ 1.5 &   9.2 & --- & ~~~~~~~ & 0.0 & 1 \\
367787 & 05:35:00.26 & -06:02:14.5 & M5.5$\pm$0.8 &  -19.4$\pm$ 5.0 & $<$ 1.3 &  11.8 & WTTS & ~~~~5~~ & 0.0 & 1 \\
368458 & 05:35:01.37 & -05:55:54.8 & M0.0$\pm$1.1 &   -4.1$\pm$ 0.9 &     0.9 &  55.8 & WTTS & 12~~~67 & 0.0 & 1 \\
368597 & 05:35:01.61 & -05:58:21.8 & M5.3$\pm$3.0 &    --- & $<$ 1.2 &   6.8 & --- & ~2~~~~~ & 0.0 & 1 \\
369203 & 05:35:02.94 & -06:05:25.2 & M0.8$\pm$0.6 &   -2.9$\pm$ 1.5 &     0.5 & 116.6 & WTTS & 1234~~7 & 0.0 & 1 \\
370047 & 05:35:04.52 & -05:55:25.7 & M5.2$\pm$0.6 &   -9.9$\pm$ 4.6 &     0.7 &  51.9 & WTTS & 1~~4~~7 & 0.0 & 1 \\
370883 & 05:35:06.09 & -05:52:39.3 & M4.6$\pm$2.6 &    1.9$\pm$13.2 &     0.4 &  20.0 & WTTS & 12~~~67 & 0.0 & 1 \\
372058 & 05:35:08.38 & -05:54:41.7 & M4.9$\pm$0.5 &   -7.8$\pm$ 0.6 &     0.6 &  66.1 & WTTS & 12~~~~7 & 0.0 & 1 \\
372451 & 05:35:09.00 & -05:58:17.1 & M0.9$\pm$1.3 &    \tablenotemark{c} &     0.8 &  32.8 & --- & 12~~~~~ & 0.0 & 1 \\
373436 & 05:35:11.11 & -06:00:21.7 & M4.6$\pm$0.6 &   -7.2$\pm$ 0.1 &     0.4 &  67.3 & WTTS & 1~34~~7 & 0.0 & 1 \\
377519 & 05:35:18.68 & -05:54:06.4 & M2.8$\pm$0.9 &    --- &     0.6 &  33.4 & --- & 12~~~~7 & 0.0 & 1 \\
378822 & 05:35:21.32 & -05:56:36.0 & M5.5$\pm$1.5 &    --- &     --- &   4.9 & --- & ~2~~~6~ & --- & 1 \\
378963 & 05:35:21.33 & -05:53:51.8 & M0.7$\pm$1.2 &    --- &     0.6 &  31.1 & --- & 12~~~~~ & 0.6 & 1 \\
378903 & 05:35:21.48 & -05:57:42.2 & M3.5$\pm$1.2 &    --- &     1.2 &  26.0 & --- & 12~~~~~ & 0.0 & 1 \\
380723 & 05:35:25.21 & -05:55:55.3 & M5.3$\pm$1.6 &    \tablenotemark{c} & $<$ 1.0 &   7.5 & --- & ~2~~~6~ & --- & 1 \\
380886 & 05:35:25.38 & -05:53:21.6 & M4.5$\pm$2.1 &    --- &     --- &   3.0 & --- & ~2~~56~ & --- & 1 \\
381182 & 05:35:26.01 & -06:01:39.8 & M3.3$\pm$0.7 &   -3.6$\pm$ 0.9 &     0.5 &  82.4 & WTTS & 1~~~~~7 & --- & 1 \\
384731 & 05:35:33.00 & -05:58:03.0 & M2.0$\pm$0.6 &   -6.1$\pm$ 0.2 &     0.5 & 137.6 & WTTS & 12~~~~~ & 0.0 & 1 \\
385017 & 05:35:33.58 & -06:02:20.4 & M3.8$\pm$0.7 &   -6.0$\pm$ 0.3 &     0.5 & 119.4 & WTTS & 1234~~7 & 0.0 & 1 \\
386326 & 05:35:36.11 & -05:56:53.3 & M4.7$\pm$0.5 &   -9.9$\pm$ 0.2 &     0.6 &  70.5 & WTTS & 12~~~~7 & 0.0 & 1 \\
387543 & 05:35:38.44 & -05:52:45.7 & M5.4$\pm$0.9 &  -42.5$\pm$ 7.1 &     0.8 &  10.6 & CTTS & 1~~~56~ & 0.4 & 1 \\
387551 & 05:35:38.45 & -05:56:23.8 & M5.8$\pm$0.7 &  -20.4$\pm$ 7.7 & $<$ 0.7 &   9.4 & --- & ~~~~~~~ & 0.1 & 1 \\
388596 & 05:35:40.56 & -05:58:26.2 & M5.1$\pm$0.8 &   -9.2$\pm$ 2.0 &     0.5 &  15.9 & WTTS & 1~~~~~~ & 0.0 & 1 \\
393298 & 05:35:49.24 & -05:59:14.3 & M4.3$\pm$0.8 &   -5.5$\pm$ 2.5 &     0.3 &  45.4 & WTTS & 1~~~~~7 & 0.0 & 1 \\
370107 & 05:35:04.80 & -06:00:20.9 & M0.4$\pm$0.6 &   -3.6$\pm$ 0.0 &     0.6 & 128.0 & WTTS & 123~~6~ & 0.0 & 1 \\
377886 & 05:35:19.16 & -06:04:37.8 & M4.3$\pm$0.7 &   -5.0$\pm$ 0.1 &     0.6 &  75.9 & WTTS & 1~3~~~~ & 0.0 & 1 \\
373101 & 05:35:10.48 & -05:52:19.1 & M4.3$\pm$1.6 &   -8.7$\pm$ 9.4 & $<$ 1.5 &   7.0 & --- & ~~~~~~~ & 0.2 & 2 \\
353575 & 05:34:37.95 & -05:46:10.7 & M4.9$\pm$1.0 &  -13.5$\pm$ 2.5 & $<$ 1.2 &  11.4 & WTTS & ~~~~~~7 & 0.0 & 2 \\
354003 & 05:34:38.56 & -05:47:35.2 & M0.8$\pm$0.7 &   -2.5$\pm$ 1.3 &     0.5 & 149.8 & WTTS & 1~~~~67 & 0.5 & 2 \\
355801 & 05:34:41.00 & -05:48:05.0 & M4.8$\pm$1.0 &    \tablenotemark{d} & $<$ 0.3 &  12.7 & --- & ~~~~~~7 & 0.0 & 2 \\
355912 & 05:34:41.18 & -05:46:11.7 & M2.2$\pm$0.6 &  -47.1$\pm$ 2.5 &     0.5 &  95.6 & CTTS & 1~~4567 & 3.3 & 2 \\
356429 & 05:34:41.99 & -05:45:00.7 & M5.4$\pm$1.0 & -109.4$\pm$15.3 &     0.8 &  10.6 & CTTS & 1~~~567 & 1.0 & 2 \\
356725 & 05:34:42.44 & -05:43:25.6 & M4.7$\pm$0.9 &  -42.2$\pm$ 4.4 &     1.3 &  14.5 & CTTS & 1~~456~ & 1.3 & 2 \\
357113 & 05:34:43.16 & -05:44:40.1 & M0.6$\pm$0.7 & -124.2$\pm$12.4 & $<$ 0.1 &  33.5 & CTTS & ~~~456~ & 1.8 & 2 \\
358931 & 05:34:45.90 & -05:48:12.4 & M3.6$\pm$1.0 &    --- &     0.3 &  30.1 & --- & 1~~4~~7 & 0.0 & 2 \\
359618 & 05:34:46.96 & -05:52:20.5 & M5.0$\pm$1.1 &    \tablenotemark{d} & $<$ 0.6 &  10.1 & --- & ~~~4~~7 & 0.0 & 2 \\
359980 & 05:34:47.54 & -05:46:30.4 & M1.3$\pm$0.7 &   -2.6$\pm$ 0.6 &     0.5 & 110.4 & WTTS & 12~~~67 & 0.2 & 2 \\
361728 & 05:34:50.01 & -05:53:04.2 & M5.0$\pm$0.9 &  -15.7$\pm$ 2.5 & $<$ 0.9 &   9.3 & WTTS & ~2~~~~7 & 0.0 & 2 \\
363704 & 05:34:53.13 & -05:47:42.6 & M4.3$\pm$1.2 &    \tablenotemark{d} & $<$ 0.3 &   9.2 & --- & ~~~~~6~ & 0.0 & 2 \\
363995 & 05:34:53.59 & -05:43:45.9 & M5.4$\pm$2.8 &    --- &     --- &   3.6 & --- & ~~~~~~~ & --- & 2 \\
364851 & 05:34:54.91 & -05:47:09.0 & M5.8$\pm$2.2 &    --- &     --- &   4.8 & --- & ~~~~~~~ & --- & 2 \\
366388 & 05:34:57.69 & -05:48:19.5 & M4.0$\pm$3.6 &    --- &     --- &   4.3 & --- & ~~~~~~~ & --- & 2 \\
366595 & 05:34:57.94 & -05:53:22.1 & M4.1$\pm$1.0 &   -7.6$\pm$ 3.1 & $<$ 1.6 &  19.9 & --- & ~~~~~~~ & --- & 2 \\
367010 & 05:34:58.86 & -05:47:06.8 & M5.2$\pm$4.0 &    --- &     --- &   2.9 & --- & ~~~~56~ & --- & 2 \\
367244 & 05:34:59.19 & -05:41:12.7 & M5.0$\pm$2.8 &    --- &     --- &   4.4 & --- & ~~~~56~ & --- & 2 \\
368529 & 05:35:01.61 & -05:54:18.9 & M6.1$\pm$3.7 &    --- &     --- &   2.9 & --- & ~2~~~~7 & --- & 2 \\
369121 & 05:35:02.79 & -05:44:43.1 & M5.2$\pm$1.0 &    \tablenotemark{d} & $<$ 1.7 &   9.6 & --- & ~~~~56~ & 0.9 & 2 \\
369185 & 05:35:03.04 & -05:45:33.4 & M3.1$\pm$2.4 &    \tablenotemark{d} & $<$ 1.7 &   5.5 & --- & ~~~45~~ & 3.9 & 2 \\
369518 & 05:35:03.51 & -05:51:59.1 & M4.9$\pm$1.6 &    \tablenotemark{d} & $<$ 1.5 &   5.7 & --- & ~~~~~~7 & 0.0 & 2 \\
370662 & 05:35:05.68 & -05:43:04.7 & M4.8$\pm$1.0 &    --- &     0.8 &  11.6 & --- & 1~~~~~7 & 0.0 & 2 \\
371190 & 05:35:06.71 & -05:58:11.4 & M3.2$\pm$1.1 &   -2.5$\pm$ 3.8 & $<$ 0.4 &  16.9 & --- & ~~~~~~~ & 0.2 & 2 \\
371902 & 05:35:08.06 & -05:53:43.5 & M1.6$\pm$0.9 &   -5.0$\pm$ 1.8 &     0.6 &  41.5 & WTTS & 12~~~67 & 0.0 & 2 \\
373179 & 05:35:10.63 & -06:01:44.8 & M3.3$\pm$1.0 &   -7.4$\pm$ 0.1 & $<$ 0.5 &  19.6 & WTTS & ~234~~7 & 0.0 & 2 \\
374326 & 05:35:12.83 & -05:55:26.4 & M4.2$\pm$1.5 &  -68.7$\pm$25.2 & $<$ 1.8 &   7.1 & CTTS & ~~~~567 & 0.1 & 2 \\
375135 & 05:35:14.44 & -05:54:26.7 & M4.3$\pm$1.5 &  -34.1$\pm$15.1 & $<$ 0.8 &   8.1 & CTTS & ~2~~567 & 0.5 & 2 \\
375411 & 05:35:14.94 & -05:56:36.2 & M4.4$\pm$1.3 &   -6.6$\pm$ 1.3 & $<$ 0.8 &  10.6 & WTTS & ~2~~~67 & 0.0 & 2 \\
376697 & 05:35:16.96 & -05:45:56.0 & M5.8$\pm$1.3 &  -40.2$\pm$16.3 & $<$ 0.5 &   8.6 & CTTS & ~~~4567 & 0.3 & 2 \\
377611 & 05:35:18.71 & -06:03:27.6 & M5.5$\pm$1.9 &  -19.9$\pm$12.9 & $<$ 1.0 &   5.0 & WTTS & ~~3~~~7 & 0.0 & 2 \\
378163 & 05:35:19.83 & -05:45:41.0 & M4.4$\pm$1.0 &  -11.3$\pm$10.4 &     0.6 &  13.9 & WTTS & 1~~~~67 & 0.0 & 2 \\
378352 & 05:35:20.22 & -05:46:51.1 & M2.7$\pm$1.0 &   -5.1$\pm$ 5.8 &     0.6 &  32.8 & WTTS & 1~~~567 & 0.0 & 2 \\
378392 & 05:35:20.30 & -05:46:40.0 & M3.0$\pm$0.9 &   -4.0$\pm$ 4.8 &     0.8 &  27.8 & WTTS & 1~~4~67 & 0.0 & 2 \\
379001 & 05:35:21.55 & -06:02:41.7 & M6.8$\pm$5.2 &    --- &     --- &   2.1 & --- & ~~~~~~~ & --- & 2 \\
379834 & 05:35:23.28 & -06:03:01.1 & M4.9$\pm$2.1 &    --- &     --- &   4.8 & --- & ~~3~~~~ & --- & 2 \\
381261 & 05:35:26.17 & -05:45:08.6 & M2.9$\pm$2.7 &   -3.0$\pm$ 4.2 &     0.5 &  12.9 & WTTS & 1~~456~ & 1.7 & 2 \\
382206 & 05:35:28.03 & -05:57:43.9 & M4.9$\pm$1.9 &    --- &     --- &   4.5 & --- & ~~~~~~~ & --- & 2 \\
383442 & 05:35:30.45 & -06:05:00.9 & M5.5$\pm$3.2 &    --- &     --- &   2.8 & --- & ~~3~5~7 & --- & 2 \\
384154 & 05:35:31.81 & -06:06:29.4 & M2.9$\pm$0.7 &   -6.4$\pm$ 0.9 &     0.1 &  41.3 & WTTS & 123~~~~ & 0.0 & 2 \\
384868 & 05:35:33.16 & -05:47:07.4 & M3.9$\pm$2.7 &    --- &     --- &   4.6 & --- & ~2~456~ & --- & 2 \\
385084 & 05:35:33.72 & -05:46:16.3 & M3.4$\pm$0.9 &   -3.3$\pm$ 3.0 &     0.6 &  30.2 & WTTS & 1~~~~~~ & 0.0 & 2 \\
385453 & 05:35:34.39 & -06:05:42.8 & M3.6$\pm$0.7 &   -7.3$\pm$ 0.4 &     0.5 &  33.7 & WTTS & 1234~~7 & 0.2 & 2 \\
389236 & 05:35:41.86 & -05:52:29.5 & M4.7$\pm$1.1 &  -15.0$\pm$ 2.4 &     0.7 &  14.7 & WTTS & 12~~~~7 & 0.0 & 2 \\
392397 & 05:35:47.69 & -05:58:06.1 & M5.2$\pm$2.7 &   -6.9$\pm$ 3.9 & $<$ 0.8 &   7.4 & WTTS & ~2~~~~7 & 0.0 & 2 \\
395473 & 05:35:53.08 & -05:59:48.0 & M3.4$\pm$1.0 &   -5.4$\pm$ 3.4 & $<$ 0.1 &  21.6 & WTTS & ~2~~~~7 & 0.0 & 2 \\
398995 & 05:35:58.98 & -05:59:08.5 & M3.0$\pm$0.9 &   -5.1$\pm$ 1.1 &     0.5 &  29.1 & WTTS & 12~4~~7 & 0.0 & 2 \\
400107 & 05:36:00.88 & -06:01:30.4 & M5.3$\pm$4.3 &    --- &     --- &   2.1 & --- & ~~~~~~~ & --- & 2 \\
372153 & 05:35:08.57 & -05:53:51.4 & M4.2$\pm$1.2 &  -14.5$\pm$ 1.7 & $<$ 2.1 &   9.9 & --- & ~~~~~~~ & 0.0 & 3 \\
373336 & 05:35:10.88 & -05:49:48.6 & M3.9$\pm$1.4 &  -10.9$\pm$ 1.9 & $<$ 1.1 &   7.0 & WTTS & ~~~~~~7 & 0.0 & 3 \\
373772 & 05:35:11.93 & -05:45:38.0 & M1.5$\pm$2.1 &    --- & $<$ 1.0 &   9.5 & --- & ~2~4~6~ & 0.7 & 3 \\
374828 & 05:35:13.77 & -05:39:10.1 & K7.1$\pm$1.0 &   -8.0$\pm$ 2.4 &     0.6 &  82.0 & WTTS & 1~~~56~ & 1.8 & 3 \\
375422 & 05:35:14.94 & -05:49:34.8 & M2.5$\pm$1.2 &   -5.2$\pm$ 0.6 &     0.7 &  19.1 & WTTS & 1~~~~~7 & 0.3 & 3 \\
375439 & 05:35:14.92 & -05:36:39.2 & M2.6$\pm$1.3 &  -21.0$\pm$11.9 &     1.3 &  12.2 & CTTS & 1~~456~ & 0.9 & 3 \\
375615 & 05:35:15.28 & -05:56:08.6 & M3.3$\pm$2.0 &    --- & $<$ 1.8 &   5.9 & --- & ~~~~~~~ & 0.0 & 3 \\
375665 & 05:35:15.31 & -05:39:56.3 & M3.3$\pm$1.1 &   -8.1$\pm$ 7.5 &     0.7 &  13.2 & WTTS & 1~~4~6~ & 0.7 & 3 \\
376797 & 05:35:17.16 & -05:41:54.0 & M4.3$\pm$0.9 &  -13.7$\pm$ 1.6 & $<$ 0.5 &  15.7 & WTTS & ~~~~56~ & 0.4 & 3 \\
377595 & 05:35:18.68 & -05:41:09.6 & M8.2$\pm$4.1 &    --- &     --- &   2.0 & --- & ~~~~56~ & --- & 3 \\
377784 & 05:35:19.04 & -05:46:16.3 & M3.0$\pm$0.8 &   -2.7$\pm$ 0.8 &     0.6 &  31.6 & WTTS & 12~~~~7 & 0.1 & 3 \\
381346 & 05:35:26.35 & -05:43:36.6 & M5.0$\pm$2.1 &    --- &     --- &   2.8 & --- & ~~~~~~~ & --- & 3 \\
383458 & 05:35:30.48 & -05:49:03.7 & M2.3$\pm$1.3 &   -8.8$\pm$ 3.2 &     0.6 &  22.3 & WTTS & 1~~456~ & 1.2 & 3 \\
385167 & 05:35:33.85 & -05:38:20.7 & M3.3$\pm$0.6 &    \tablenotemark{c} & $<$ 0.6 &  20.1 & --- & ~~~~~6~ & 0.0 & 3 \\
386237 & 05:35:35.95 & -05:38:42.8 & M3.7$\pm$0.8 &  -11.7$\pm$ 2.2 & $<$ 0.1 &  21.7 & WTTS & ~~~4567 & 0.2 & 3 \\
389554 & 05:35:42.51 & -05:40:57.8 & M4.9$\pm$1.7 &    --- &     --- &   4.9 & --- & ~~~~~~~ & --- & 3 \\
393255 & 05:35:49.15 & -05:56:18.3 & M3.2$\pm$2.7 &   -6.9$\pm$ 2.2 & $<$ 0.7 &   7.4 & WTTS & ~2~~~~7 & 0.0 & 3 \\
396419 & 05:35:54.68 & -05:54:36.7 & M3.6$\pm$2.3 &  -11.4$\pm$ 2.1 & $<$ 0.8 &   7.0 & WTTS & ~~~~~~7 & 0.0 & 3 \\
397232 & 05:35:55.97 & -05:42:26.2 & M3.3$\pm$0.8 &   -4.3$\pm$ 1.1 & $<$ 0.8 &  24.4 & WTTS & ~~~4~~~ & 0.2 & 3 \\
397345 & 05:35:56.18 & -05:52:28.5 & M2.1$\pm$1.2 &   -9.7$\pm$ 2.7 & $<$ 0.5 &  20.6 & WTTS & ~2~4~~7 & 0.3 & 3 \\
375378 & 05:35:14.83 & -05:38:05.6 & K7.7$\pm$1.8 &   -8.3$\pm$10.4 & $<$ 0.5 &  17.0 & WTTS & ~~~~56~ & --- & 3 \\
361116 & 05:34:49.11 & -05:46:05.1 & M1.0$\pm$2.0 &  -58.5$\pm$ 4.2 & $<$ 1.4 &   8.2 & CTTS & ~2~456~ & 0.8 & 4 \\
364923 & 05:34:54.91 & -05:46:44.1 & K6.4$\pm$3.7 &   -6.4$\pm$ 3.1 &     0.7 &  14.4 & WTTS & 12~456~ & 3.1 & 4 \\
366774 & 05:34:58.21 & -05:49:34.8 & M3.0$\pm$3.8 &    --- &     --- &   4.2 & --- & ~~~~~~~ & --- & 4 \\
376660 & 05:35:16.82 & -05:39:17.1 & M8.9$\pm$3.2 &    --- &     --- &   1.4 & --- & ~~~4~~~ & --- & 4 \\
336633 & 05:34:13.47 & -05:55:41.7 & M2.7$\pm$2.1 &   -8.2$\pm$ 0.6 &     0.9 &  14.8 & WTTS & 1~~45~7 & 1.0 & 4 \\
338495 & 05:34:16.20 & -05:56:38.1 & M8.8$\pm$4.4 &    --- &     --- &   1.3 & --- & ~~~~~~~ & --- & 4 \\
340293 & 05:34:19.09 & -05:59:29.5 & M3.2$\pm$2.0 &  -14.3$\pm$ 1.5 & $<$ 0.5 &   5.1 & --- & ~~~~~~~ & --- & 4 \\
345759 & 05:34:27.16 & -05:48:03.7 & M4.5$\pm$2.9 &    --- &     --- &   4.0 & --- & ~~~~~~~ & --- & 4 \\
346259 & 05:34:27.87 & -05:56:28.1 & M4.6$\pm$3.6 &    --- &     --- &   4.5 & --- & ~~~~~~7 & --- & 4 \\
347500 & 05:34:29.61 & -05:47:24.7 & M1.9$\pm$2.1 &  -10.4$\pm$ 4.5 & $<$ 1.6 &  11.2 & CTTS & ~2~4567 & 0.5 & 4 \\
350678 & 05:34:33.97 & -05:48:24.9 & M4.4$\pm$2.9 &    --- &     --- &   3.8 & --- & ~~~4567 & --- & 4 \\
350838 & 05:34:34.26 & -06:02:09.7 & M2.5$\pm$1.5 &   -8.1$\pm$ 0.5 & $<$ 1.5 &   9.6 & WTTS & ~23~~~7 & 0.0 & 4 \\
353158 & 05:34:37.45 & -05:54:27.0 & M8.6$\pm$3.6 &    --- &     --- &   1.8 & --- & ~~~~~~~ & --- & 4 \\
353209 & 05:34:37.51 & -05:57:34.3 & M6.2$\pm$4.5 &    --- &     --- &   2.0 & --- & ~~~~~~~ & --- & 4 \\
355864 & 05:34:41.10 & -05:47:39.5 & M9.5$\pm$3.1 &    --- &     --- &   1.7 & --- & ~~~4~~7 & --- & 4 \\
355973 & 05:34:41.26 & -05:52:45.6 & M6.4$\pm$3.9 &    --- &     --- &   2.8 & --- & ~~~~~~7 & --- & 4 \\
357985 & 05:34:44.45 & -05:56:15.0 & M1.5$\pm$2.1 &    --- & $<$ 0.4 &   7.5 & --- & ~2~4~6~ & 0.1 & 4 \\
361467 & 05:34:49.81 & -05:44:55.1 & M6.6$\pm$3.5 &    --- &     --- &   2.0 & --- & ~~~~~~~ & --- & 4 \\
362886 & 05:34:51.75 & -05:39:24.1 & M3.0$\pm$1.8 &  -14.5$\pm$ 9.2 & $<$ 0.2 &   9.8 & WTTS & ~~~456~ & 1.1 & 4 \\
364118 & 05:34:53.69 & -05:45:11.3 & M8.0$\pm$5.0 &    --- &     --- &   1.6 & --- & ~~~4~~~ & --- & 4 \\
365786 & 05:34:56.40 & -05:38:05.3 & M1.9$\pm$3.0 &  -17.3$\pm$ 3.3 & $<$ 0.7 &   6.5 & CTTS & ~~~~~67 & 0.0 & 4 \\
366163 & 05:34:57.24 & -05:42:02.9 & M0.3$\pm$3.9 &   -6.8$\pm$ 6.3 & $<$ 0.9 &   8.8 & WTTS & ~~~4~6~ & 0.2 & 4 \\
366573 & 05:34:58.06 & -05:53:43.0 & M4.8$\pm$2.2 &    --- &     --- &   3.6 & --- & ~~~~~~~ & --- & 4 \\
369855 & 05:35:04.07 & -05:48:54.2 & M4.7$\pm$4.0 &    --- &     --- &   3.1 & --- & ~~~4~~~ & --- & 4 \\
370374 & 05:35:05.07 & -05:36:43.9 & M6.5$\pm$4.9 &    --- &     --- &   3.5 & --- & ~~~456~ & --- & 4 \\
382601 & 05:35:28.78 & -05:41:34.1 & M3.2$\pm$1.9 &  -12.6$\pm$ 2.1 & $<$ 0.3 &   5.8 & --- & ~~~~~~~ & 0.3 & 4 \\
382894 & 05:35:29.31 & -05:45:38.2 & M5.3$\pm$4.0 &    --- & $<$ 1.3 &   6.4 & --- & ~~~4~6~ & 0.0 & 4 \\
\enddata
\tablenotetext{a}{Reported only for sources with SNR$>$5}
\tablenotetext{b}{1: This work, 2: \citet{2013Pillitteri}, 3: \citet{2012Hsu}, 4: \citet{2008furesz}, 5: \citet{2012megeath}, 6: \citet{2012Morales}, 7: \citet{2017fang}}
\tablenotetext{c}{Line has either fallen onto the chip gap or off the edge of the field}
\tablenotetext{d}{Poor nebular line subtraction}
\tablenotetext{}{Full version of the table will be available in the online text.}
\end{deluxetable*}

\end{document}